\documentclass[prd,twocolumn,showpacs,nofootinbib,tighten]{revtex4}

\usepackage{amsmath,amssymb,amsfonts}
\usepackage{mathrsfs}
\usepackage{graphicx}
\usepackage{subfigure}
\usepackage[]{natbib}
\usepackage{enumerate}
\usepackage{color}

\usepackage[plainpages=false, colorlinks=true, anchorcolor=blue, linkcolor=blue, citecolor=blue, bookmarks=false]{hyperref}

\def\be{\begin{equation}}
\def\ee{\end{equation}}
\newcommand{\lp}{\left(}
\newcommand{\rp}{\right)}
\newcommand{\bb}{\begin{bmatrix}}
\newcommand{\eb}{\end{bmatrix}}

\newcommand{\msJ}{\mathcal{J}}

\begin{document}

\title{A Trans-dimensional Bayesian Approach to Pulsar Timing Noise Analysis}

\author{J.~A.~Ellis}
\email{Justin.A.Ellis@jpl.nasa.gov}
\affiliation{Jet Propulsion Laboratory, California Institute of Technology, 4800 Oak Grove Drive, Pasadena, CA 91109, USA}
\altaffiliation{Einstein Fellow}

\author{N.~J.~Cornish}
\affiliation{Department of Physics, Montana State University, Bozeman, MT 59717, USA}

\date{\today}


\begin{abstract}
The modeling of intrinsic noise in pulsar timing residual data is of crucial importance for Gravitational Wave (GW) detection and pulsar timing (astro)physics in general. The noise budget in pulsars is a collection of several well studied effects including radiometer noise, pulse-phase jitter noise, dispersion measure (DM) variations, and low frequency spin noise. However, as pulsar timing data continues to improve,  non-stationary and non-powerlaw noise terms are beginning to manifest which are not well modeled by current noise analysis techniques. In this work we use a trans-dimensional approach to model these non-stationary and non-powerlaw effects through the use of a wavelet basis and an interpolation based adaptive spectral modeling. In both cases, the number of wavelets and the number of control points in the interpolated spectrum are free parameters that are constrained by the data and then marginalized over in the final inferences, thus fully incorporating our ignorance of the noise model. We show that these new methods outperform standard techniques when non-stationary and non-powerlaw noise is present. We also show that these methods return results consistent with the standard analyses when no such signals are present.
\end{abstract}


\maketitle


\section{Introduction}
\label{sec:intro}

Pulsar timing has been used as a tool to probe important aspects of astrophysics and fundamental physics including the indirect detection of gravitational waves (GWs) \cite{ht75}, the precise determination of neutron star masses \citep{dpr+10}, precise tests of General Relativity and alternative theories of gravity \cite{s03}, constraints on equation-of-state physics, and several other areas \cite{l08}. Furthermore, the goal of modern PTAs  is the direct detection of low-frequency gravitational waves (GWs) which could usher in a new era of astronomy and astrophysics \cite{l15}.

As pulsar timing becomes more precise, increasingly robust and sophisticated noise modeling will be needed in order to make (astro)physical inferences from the data. In the last several years noise modeling has become an active area of research within the pulsar community \citep{cs10, sc10, vhl13, kcs+13, lah+14, css15, lcc+15, lcc+15b} and has proven crucial to current timing and GW detection efforts \citep{abb+14, abb+15, zsd+15, ltm+15, abb+15b, bps+16, rhc+16}. However, even with current sophisticated methods there are still several areas in which our current noise models are inadequate:
\begin{enumerate}
\item We currently have no model for non-white non-stationary noise features which could result from a transient instrumental or astrophysical source. 
\item We have no robust way of determining the validity or the goodness of fit of our low-frequency red-noise modeling.
\end{enumerate}

In this work we seek to address these two problems by using a trans-dimensional approach in which a variable number of wavelets model the non-white, non-stationary features and a variable number of interpolation control points determine the spectrum of the stationary red noise.  Inferences on any parameters of interest (i.e. timing model, GW parameters, etc.) are then marginalized over all possible models (i.e. marginalized over the number of wavelets and/or control points) thereby incorporating our uncertainty in the noise \emph{model} itself as opposed to the standard practice of choosing \emph{one} model and incorporating the uncertainty from the \emph{parameters} within that model. There is precedent for this type of analysis in other GW experiments, namely the \texttt{BayesWave} \cite{cl15} and \texttt{BayesLine} \cite{lc15} algorithms used in LIGO/Virgo analyses, both of which serve as a starting point for this work.

The paper is organized as follows: In Section \ref{sec:da} we review current Bayesian data analysis techniques use in pulsar timing. In Sections \ref{sec:nstat} and \ref{sec:spectrum} we introduce the non-stationary and adaptive spectral modeling techniques. In Section \ref{sec:techniques} we review the Bayesian algorithms used to implement our trans-dimensional  models. In Section \ref{sec:results} we test our analysis on simulated data and compare against traditional techniques. Finally we conclude in Section \ref{sec:conclude}.

\section{Standard Pulsar Timing Data Analysis}
\label{sec:da}

The basic data analysis techniques used in this work are very similar to that presented in \cite{abb+15,abb+15b}. Here we will briefly review the formalism and set up the likelihood function for this work.

\subsection{Likelihood} \label{sec:likelihood}

Pulsar timing data consists of a set of times-of-arrival (TOAs) or the radio pulses and a timing model. Our $N_{\rm TOA}$ timing residuals (difference of measured and modeled TOAs) for a single pulsar $\delta\mathbf{t}$ can be broken into individual components as follows:
\begin{equation}
\delta\mathbf{t} = M\boldsymbol{\epsilon} + F\mathbf{a} + U\mathbf{j}  + \mathbf{s} + \mathbf{n} .
\label{eq:timingresiduals}
\end{equation}
The term $M\boldsymbol{\epsilon}$ describes inaccuracies in the subtraction of
the timing model, where $M$ is called the timing model design matrix, and
$\boldsymbol{\epsilon}$ is a vector of small timing model parameter offsets. The
term $F\mathbf{a}$ describes all low-frequency signals, including low-frequency
(``red'') noise,
with a limited number of Fourier coefficients $\mathbf{a}$. Our harmonics are chosen as integer multiples of the harmonic base frequency
$1/T$, with $T$ the length of our dataset . The matrix $F$ then has alternating sine and cosine
functions.  The term $U\mathbf{j}$ describes noise that is
fully correlated for simultaneous observations at different observing
frequencies, but fully uncorrelated in time. The matrix $U$ is an $N_{\rm
TOA}\times N_{\rm epoch}$ matrix that maps the $N_{\rm epoch}$ observation sessions to the $N_{\rm
TOA}$ TOAs.
The vector $\mathbf{j}$ describes the white noise per
observation session that is fully correlated across all observing frequencies.
The term $\mathbf{s}$ is simply any other deterministic feature that we choose to include in our model.
The last term, $\mathbf{n}$, describes Gaussian white noise that is assumed to remain in the data, after correcting for all known systematics. The white noise is assumed to be an uncorrelated and Gaussian as is described in \cite{abb+15, abb+15b}. 

We define the noise-mitigated timing residuals as
\begin{equation}
    \mathbf{r} = \delta\mathbf{t}-M\mathbf{\epsilon}-F\mathbf{a}-U\mathbf{j}-\mathbf{s},
\end{equation}
where $\mathbf{r}$ is our best approximation of $\mathbf{n}$, given our knowledge of all the noise and signal parameters. The likelihood can now be written 
\begin{equation}
    p(\delta\mathbf{t}|\boldsymbol{\epsilon}, \mathbf{a}, \mathbf{j},\mathbf{\phi}) = \frac{\exp\left( -\frac{1}{2}\mathbf{r}^TN^{-1}\mathbf{r} \right)}{\sqrt{\det(2\pi N)}},
    \label{eq:expandedlikelihood}
\end{equation}
where $N$ is the covariance matrix of the white noise. We have collectively denoted all parameters not directly represented by $\boldsymbol{\epsilon}$, $\mathbf{a}$, and $\mathbf{j}$ as $\mathbf{\phi}$. 
We group the linear signals as follows:
\begin{equation}
    T = \bb M  & F  & U \eb, \quad
    \mathbf{b} = \bb \boldsymbol{\epsilon} \\ \mathbf{a} \\ \mathbf{j} \eb,
\end{equation}
which allows us to elegantly place a Gaussian prior on the coefficients of these
random processes. The prior covariance is:
\begin{equation}
    B = \bb \infty & 0 & 0 \\ 0 & \varphi & 0 \\ 0 & 0 & \msJ \eb,
\end{equation}
resulting in a prior:
\begin{equation}
    p(\mathbf{b}|\boldsymbol{\phi}) = \frac{\exp\left(
    -\frac{1}{2}\mathbf{b}^{T}B^{-1}\mathbf{b} \right)}{\sqrt{\det(2\pi B)}},
    \label{eq:expandedprior}
\end{equation}
where $\infty$ is a diagonal matrix of infinities, which effectively means we
have a uniform unconstrained prior on the timing model parameters
$\mathbf{\epsilon}$. As described in \cite{abb+15}, this representation allows
us to analytically marginalize Eq. \eqref{eq:expandedlikelihood} times Eq. \eqref{eq:expandedprior}
over the waveform coefficients $\mathbf{b}$ resulting in the marginalized likelihood
\begin{equation}
    p(\delta\mathbf{t}|\boldsymbol{\phi}) = \frac{\exp\left(
    -\frac{1}{2}(\delta\mathbf{t}-\mathbf{s})^TC^{-1}(\delta\mathbf{t}-\mathbf{s})
    \right)}{\sqrt{\det(2\pi C)}},
    \label{eq:reducedlikelihood}
\end{equation}
with $C = N + TBT^{T}$. The Woodbury matrix identity \citep{w50} can be used to
evaluate Eq.~\eqref{eq:reducedlikelihood} efficiently. Notice that this is nearly identical
to the likelihood of \cite{abb+15,abb+15b} except that now we are including the additional
non-linear deterministic feature $\mathbf{s}$.

The parameters that describe $B$ are the hyperparameters $\boldsymbol{\phi}$.
The
hyperparameters of the diagonal matrix $\msJ$ are the per-backend {\sc TEMPO2} ECORR
parameters. The matrix $\varphi$ represents the spectrum of the low-frequency noise
 Denoting frequency bin with $(i,j)$, we can write:
\begin{equation}
    [\varphi]_{ij}=\frac{1}{T}S_{n}(f_{i})\delta_{ij},
\end{equation}
where $S_{n}(f_{i})$ is the noise power spectral density (PSD) at frequency $f_{i}$.

\subsection{Bayesian analysis} 
\label{sec:bayes}

Bayesian inference is a method of statistical inference in which Bayes rule of conditional probabilities is used to update one's knowledge as observations are acquired. Given a model $\mathcal{M}$, model parameters $\Theta$, and observations $d$, we write Bayes rule as:
\begin{equation}
p(\Theta|d,\mathcal{M}) = \frac{p(d|\Theta,\mathcal{M})p(\Theta|\mathcal{M})}{p(d|\mathcal{M})}
    \label{eq:bayes}
\end{equation}
where $p(\Theta|d,\mathcal{M})$ is the posterior probability (probability of obtaining the model parameters $\Theta$ given the data $d$ and model $\mathcal{M}$), $p(d|\Theta,\mathcal{M})$ is the likelihood function (probability of obtaining the data $d$ given the model parameters $\Theta$ and model $\mathcal{M}$), $p(\Theta|\mathcal{M})$ is the prior probability of the model parameters $\Theta$ given a model $\mathcal{M}$, and $p(d)$ is the marginal likelihood or the Bayesian evidence, usually denoted by $\mathcal{Z}$.

The left-hand side of Eq.~\eqref{eq:bayes} can be regarded as the ``output'' of the Bayesian analysis, and the right-hand side is the ``input'' modulo the evidence term. Indeed, provided we have a generative model of our observations (meaning we can simulate data, given the model parameters), we know the likelihood and prior. However, for parameter estimation we would like to know the posterior, and for model selection we need the evidence.

For parameter estimation, the evidence is usually ignored, and one can use $p(d|\Theta,\mathcal{M})p(\Theta|\mathcal{M})$ directly to map out the posterior distribution (up to a normalizing constant) and  to estimate confidence intervals. When $\Theta$ is higher-dimensional, Monte-Carlo sampling methods are typically
used to perform this multi-dimensional integral. We use Markov Chain Monte Carlo methods in this work to sample the posterior distribution.

Model selection between two models $\mathcal{M}_0$ and $\mathcal{M}_1$ can be carried by calculating the ``Bayes factor'': the ratio between the evidence for the two models. Assuming we have a prior degree of belief of how likely the two model are ($p(\mathcal{M}_0)$ and $p(\mathcal{M}_1)$), we can write:
\begin{equation}
    O = \frac{p(\mathcal{M}_1|d)}{p(\mathcal{M}_0|d)} =
    \mathcal{B}_{10}(d) \frac{p(\mathcal{M}_1)}{p(\mathcal{M}_0)},
    \label{eq:oddsratio}
\end{equation}
where $\mathcal{B}_{10}(d) \equiv \mathcal{Z}_1/\mathcal{Z}_0$ is the Bayes factor, and $O$ is the odds ratio. The odds ratio can be obtained by calculating the evidence $\mathcal{Z}$ for each model separately (e.g. with Nested Sampling or thermodynamic integration), or by calculating the Bayes factor $\mathcal{B}$ between two models directly (e.g. with trans-dimensional markov chain Monte Carlo methods).

\section{Non-stationary Noise Model}
\label{sec:nstat}

To model non-stationary noise events we use a non-orthogonal frame of Morlet-Gabor wavelets with a functional form 
\be
\label{eq:wavelet}
\Psi(t;A,f_{0},Q,t_{0}, \phi_{0}) = Ae^{-(t-t_{0})^{2}/\tau^{2}}\cos(2\pi f_{0}(t-t_{0})+\phi_{0}),
\ee
where $\tau=Q/(2\pi f_{0})$ is the width of the gaussian envelope, $A$ is the amplitude, $t_{0}$ and $f_{0}$ are the central time and frequency of the wavelet and $\phi_{0}$ is a phase offset. This frame is simply a matter of choice and could be replaced by any other basis, such as shapelets \citep{r03}. However, this basis is well suited to modeling non-stationary transient events in our residual data because they are localized in both time and frequency and their shape in time-frequency space can adapt to fit the signal. Thus, to model non-stationary noise events we use a collection of independent wavelets where the number of wavelets, $N_{\rm wave}$, is a free parameter that must be constrained by the data. In principle, this analysis is similar to that used in \cite{lah15, ls15} where shapelets are used to model the pulse profile; however, in this case the number of wavelets is free to vary, where in the shapelet analysis separate runs must be done in order to find the number of shapelets with the highest Bayes factor. 

As can be seen from Eq. \eqref{eq:wavelet}, the Morlet-Gabor wavelet is a function of five parameters, thus this analysis includes an additional $5N_{\rm wave}$ over the standard gaussian noise model. Our RJMCMC implementation will therefore balance the goodness-of-fit for the likelihood with the model complexity (i.e., prior volume) of the additional wavelets. Getting the correct balance of the two aforementioned effects is a conceptually and computationally difficult problem that depends heavily on the choice of priors on the wavelet parameters and on the number of wavelets.

\subsection{Wavelet Priors}

As we will see, the priors on the wavelet amplitude are quite important in balancing goodness-of-fit with the model complexity.   While the priors on the other parameters will likely have some effect on our trans-dimensional model, for this work we use uniform priors on $\log_{10}(f_{0})$, $Q$, $t_{0}$, and $\phi_{0}$ with ranges $\log_{10}(f_{0})\in [\log_{10}(3/T), \log_{10}(1/4\Delta t)]$, with $\Delta t=2$ weeks$^{-1}$, $Q\in[0.5, 40]$, $t_{0}\in [t_{\rm start}, t_{\rm stop}]$, where $t_{\rm start}$, and $t_{\rm stop}$ are the start and end times of the residuals time-series, and $\phi_{0}\in[0,2\pi]$. Here we will only focus on choosing a good amplitude prior that in effect will limit the number of wavelets used in the RJMCMC. When doing standard fixed dimension noise analysis it is standard to use a log-uniform prior on the amplitude of deterministic sources unless we have some a-priori information; however, in this trans-dimensional model a log-uniform prior will lead to many wavelets being used that essentially just explore the prior and contribute little to the signal being modeled. As is used in the \texttt{BayesWave} algorithm \citep{cl15},  a more reasonable prior is one that is based on the SNR of the wavelet
\be
\rho^{2} = (\Psi|\Psi) = \Psi^{T}C^{-1}\Psi,
\ee
where $\Psi$ is the wavelet waveform and $C$ is the full covariance matrix from Eq. \eqref{eq:reducedlikelihood}. In practice, computing the SNR as defined above carries nearly the same computational cost as that of computing the likelihood. Furthermore, to use the SNR in a prior we will need to compute it for \emph{every} wavelet used in the model. Therefore, it is computationally undesirable to use the full SNR but to instead use an SNR proxy that is computationally efficient. For this work we choose a simple white noise inner product
\be
\bar{\rho}^{2} = \Psi^{T}N^{-1}\Psi = A^{2}\bar{\Psi}^{T}N^{-1}\bar{\Psi},
\label{eq:rhoamp}
\ee
where $N$ is the diagonal white noise matrix and $\bar{\Psi}$ is the amplitude-free waveform. In the simulations presented later in this paper we have found that this overestimates the true SNR by a factor of two in many cases; however this proxy is very fast to compute and scales similarly with the other wavelet parameters as the true SNR. Using this SNR proxy, we parameterized the signal in terms of $\bar{\rho}$ as 
\be
\Psi(t) = \frac{\bar{\rho}}{(\bar{\Psi}^{T}N^{-1}\bar{\Psi})^{1/2}}\bar{\Psi}(t),
\ee
where $\bar{\rho}$ is now the free parameter. As was done in \cite{cl15} we would like to use a parameterized prior on $\bar{\rho}$ that peaks at some characteristically detectable value and slowly decreases as $\bar{\rho}$ gets large; however, since we are using an SNR-proxy the choice of the peak in the distribution is problem specific and is very difficult to determine without some a-priori knowledge. For this reason we have chosen a simple uniform prior of $\bar{\rho}\in[0,100]$ for this work. Finally we choose a uniform prior on the number of active wavelets, $N_{\rm wave}\in[0,30]$. We have experimented with various exponential priors on $N_{\rm wave}$ but we have found that these have little effect and the choice of wavelet amplitude prior is the dominant factor in constructing a parsimonious model. Further consideration of these prior choices will be addressed in a future work. 

\section{Adaptive Spectral Modeling}
\label{sec:spectrum}

As mentioned in Section \ref{sec:intro}, current noise analysis methods rely on specifying a model a-priori and then performing Bayesian model selection to determine the best model. However, in many cases several different spectral models seem to describe the data equally well. This uncertainty in the model and the computational complexity of performing several different runs to determine the Bayesian evidence has led us to develop a modified version of the \texttt{BayesLine} algorithm \citep{lc15} to model the red noise in our residual data. Here we model the power spectral density (PSD)\footnote{When we refer to the PSD here we are referring only to the red-noise component and not the white component of the spectrum.} with linear interpolation in the log-log space of the the PSD and frequency. The parameters of the model are the number, $N_{c}$, and the location in the frequency-PSD space $\{S_{n}(f_{i}), f_{i}\}$ of each control point in the linear interpolation. Thus the logarithm of the PSD at frequency $f_{i}$ is modeled by 
\be
\begin{split}
\log\lp S_{n}(f_{i})\rp &= \log(S_{n}(f_{i-1})) + \\
&\frac{\log\lp S_{n}(f_{i+1})/S_{n}(f_{i-1}) \rp}{\log(f_{i+1}/f_{i-1})}\log(f_{i}/f_{i-1})
\end{split}
\ee
for $i\in [0,N_{c}]$. We have chosen to carry this interpolation out in the log frequency-PSD space in order to facilitate numerical stability and in order to fit the fewest parameters. In the course of this analysis we have tried other interpolation methods, including cubic splines as are used in \texttt{BayesLine} \citep{lc15} in both log and linear frequency-PSD space. It was found that our pulsar timing residual data does not have enough frequency evolution to warrant a more sophisticated interpolation technique. We note that even though we may not have a free control point at every frequency we do in fact include all frequencies in the analysis differentiating this method from those of \cite{lah+14} in which various combinations of frequencies were used in order to determine the so-called ``optimal'' model.

In principle, one could allow the frequencies at which we place the control points to be free parameters but this could lead to many ordering and double counting problems. Instead we place the frequencies on a regularly spaced grid with $\Delta f=1/T$ in the range $[1/T, n_{f}/T]$, where $n_{f}$ is the number of frequencies. In our case we use $n_{f}=50$ unless stated otherwise. We assume uniform priors on both the logarithm of the PSD amplitude, $S_{n}(f_{i})$, and on the number of active control points, $N_{c}$ The control points at the lowest and highest frequencies are always active, thus the simplest spectrum is parameterized by two parameters in this setup. Therefore, this analysis encapsulates both the standard power-law and free spectrum analyses which are parameterized by 2 and $n_{f}$ parameters, respectively. In practice, during the RJMCMC, control points can be subtracted and added with varying amplitudes. This means that the data decide the shape and complexity of the spectrum instead of prescribing a model a-priori. 

\section{Bayesian Techniques}
\label{sec:techniques}

\subsection{Reverse Jump MCMC}

The main driver of this trans-dimensional noise modeling technique is the RJMCMC algorithm, a variant of traditional MCMC where the model itself is included the explored parameter space.  Traditionally RJMCMC has been used as a model selection tool by computing the Bayes factor of competing models as the ratio of the iterations spent in each model. In our analysis it is used more as a tool to perform a type of Bayesian model averaging. In other words, we are more interested in the ability of the RJMCMC to map out the trans-dimensional distribution and provide marginalized distributions over the full model-space than we are interested in choosing a \emph{specific} model via the resulting Bayes factors of the RJMCMC. 

In the RJMCMC framework we must include a separate Metropolis Hasting step that proposes the trans-dimensional move. The parameters $\Theta_{i}$ of model $\mathcal{M}_{i}$ are drawn from the proposal distribution $q(\Theta_{i}|\mathcal{M}_{i})$. Since our model is nested (i.e. there are shared parameters between models), we hold all other parameters fixed while the new parameters are drawn from the proposal distribution. The criteria for accepting this trans-dimensional jump is the Hastings ratio
\be
\mathcal{H}_{\mathcal{M}_{i}\rightarrow\mathcal{M}_{j}}=\frac{p(d|\Theta_{j},\mathcal{M}_{j})p(\Theta_{j}|\mathcal{M}_{j})q(\Theta_{j}|\mathcal{M}_{i})}{p(d|\Theta_{i},\mathcal{M}_{i})p(\Theta_{i}|\mathcal{M}_{i})q(\Theta_{j}|\mathcal{M}_{i})}|J_{ij}|,
\ee
where the Jacobian $|J_{ij}|$ accounts for any change in dimension across models. However, if our jump proposals yield parameter values directly instead of a set of random numbers that are then used to determine the new model parameters, then the Jacobian can be neglected. Choosing a good proposal distribution is notoriously difficult and is the main drawback in using RJMCMCs. A simple choice for a RJ proposal distribution is a random draw from prior. If the models are nested and the difference in dimension between models is relatively low then this prior draw usually provided adequate mixing. 

As such, we only use prior draws for our RJ proposals in this work, for both the wavelet parameters and interpolation control points. Furthermore, we always propose to either add or subtract one component (i.e., one wavelet or one control point) and never more. In principle we could choose to add several wavelets or control points at one time but it is unlikely to be accepted. To add a new wavelet, we simply draw all 5 wavelet parameters from the prior (using either the SNR or the log-uniform prior for the amplitude). To subtract a wavelet, we randomly choose between the active wavelets at that iteration. The technique is very similar for control points. To add a new control point, we first randomly choose a frequency at which to place the control point and then draw from the log-uniform prior to choose the amplitude. To subtract a control point we randomly choose one from the active control points. When modeling both wavelets and control points simultaneously, we always hold one fixed and jump in the other, that is, we never propose to add or subtract wavelets and control points simultaneously. Of course, more complicated and more efficient proposals could be made for both the wavelet and spectral models and this will be addressed in the future; however, for this work we find that these random uniform draws provide quite adequate mixing.

\subsection{Parallel Tempering}

Although uniform trans-dimensional proposals provide adequate mixing, the trans-dimensional acceptance rates are quite low (typically $<$ 6\% for the wavelet model). This means that we have to run for many iterations in order to adequately explore the full trans-dimensional model space. In order to ameliorate this problem we use a parallel tempering method introduced in \cite{cl15} that allows different chains to be in different models, thus jumps between different temperature chains are also trans-dimensional jumps. These parallel tempering moves \emph{greatly} enhance the inter- and intra-model mixing.  Currently we use a geometric temperature spacing that is tuned to give $\sim$ 40\% acceptance for temperature swaps to the coldest chain. In future versions  of the code we will incorporate a more sophisticated adaptive temperature spacing as was introduced in \cite{vfm15}.

\section{Results}
\label{sec:results}

Here we test our non-stationary and adaptive noise modeling techniques. In all cases we have used the time stamps and timing model for PSR B1855$+$09 from the NANOGrav 9-year data release \citep{abb+15} and inject white gaussian noise consistent with the TOA uncertainties using \textsc{libstempo}\footnote{\url{https://github.com/vallis/libstempo}}. We use this as the base data set because it has all of the features of real pulsar timing data including a large gap, heteroscedastic TOA uncertainties, a full relativistic binary timing model, and time varying DM variations. We then take this base data set and inject a white noise burst, a realization of a gaussian process described by some power spectral density, or both, in order to test our modeling techniques. We then compare the results with those of standard noise analyses.

\subsection{BayesWavePTA}

\begin{figure}
\includegraphics[width=\columnwidth]{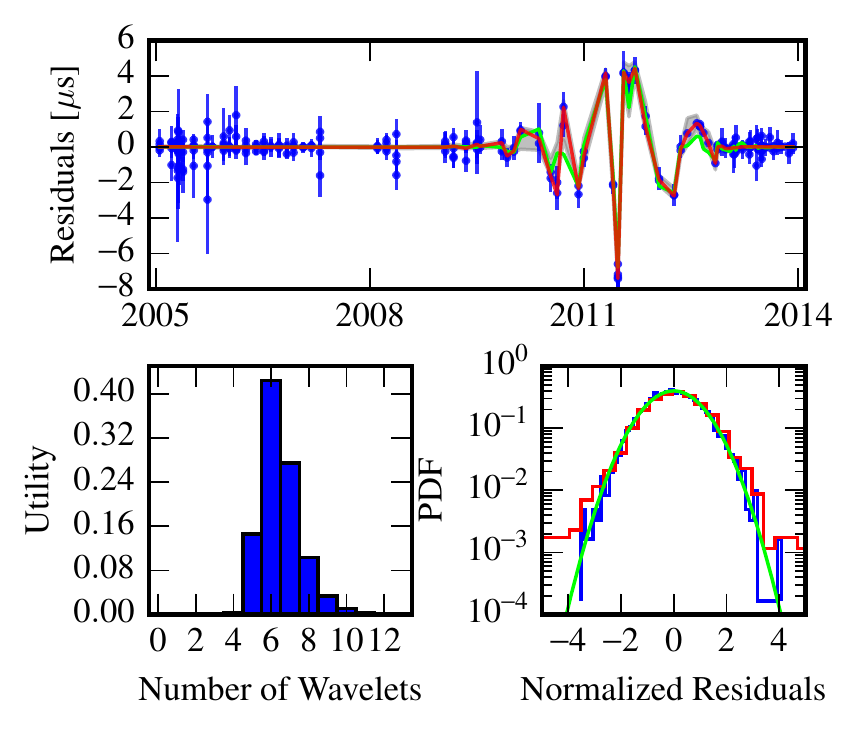} \\ 
\caption{Summary of RJMCMC results from the uniform SNR prior run. In the top panels the daily-averaged residuals are plotted in blue, the injected waveform in green, the MAP waveform (chosen from the model with the highest Bayes factor) in red, and the 90\% uncertainty on the waveform (marginalized over all models) in gray. The bottom left plot we show the utility as a function of number of wavelets. In the bottom right, we plot the histogram of the normalized residuals (normalized by the TOA uncertainties) when including (blue) and not including (red) the MAP wavelet model. The green curve is a zero-mean unit variance Gaussian distribution.}
\label{fig:bayeswavesnr} 
\end{figure}

\begin{figure}
\includegraphics[width=\columnwidth]{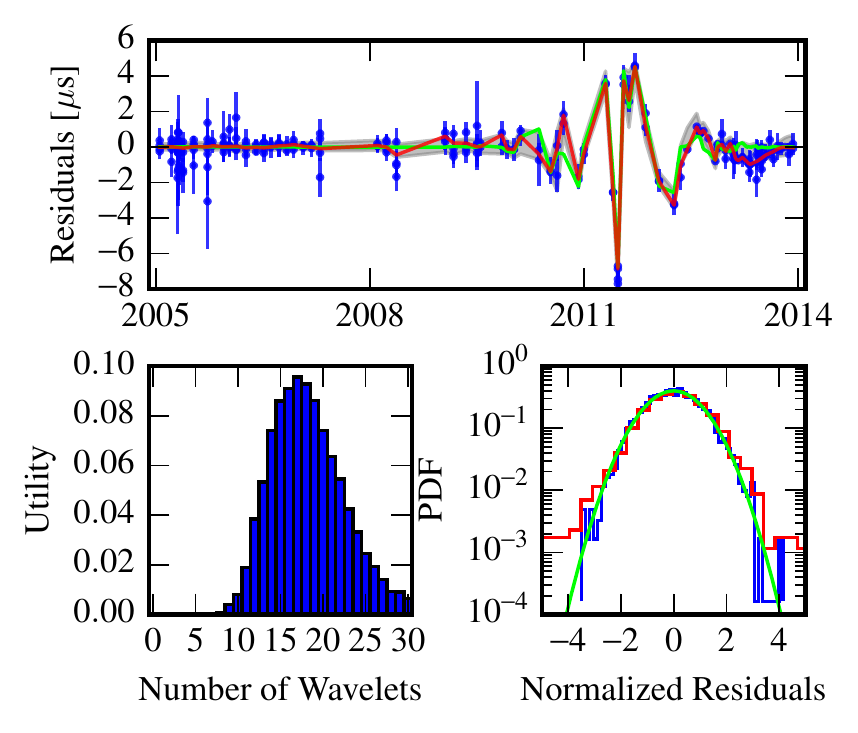} \\ 
\caption{Same as Figure \ref{fig:bayeswavesnr} but for the log-uniform wavelet amplitude RJMCMC run.}
\label{fig:bayeswavelog} 
\end{figure}

Here we test our non-stationary noise modeling techniques described in Section \ref{sec:nstat}. As mentioned above we test this technique on a simulated dataset of pulsar B1855$+$09 that contains an injected white noise burst with amplitude of $447$\,ns and a width of 250\,days. A white noise burst is simply a realization of noise from a white PSD with a gaussian window that is localized in time.  We have analyzed these data with the non-stationary model using both a uniform SNR and a uniform log-amplitude prior. In both cases we allow up to 30 wavelets with a uniform prior on the number of wavelets. The results of this analysis are shown in Figures \ref{fig:bayeswavesnr} and \ref{fig:bayeswavelog}.

In the top panels of Figures \ref{fig:bayeswavesnr} and \ref{fig:bayeswavelog} we have the daily-averaged residuals plotted in blue, the injected waveform in green, the MAP waveform (chosen from the model with the highest Bayes factor) in red, and the 90\% uncertainty on the waveform (marginalized over all models) in gray. The bottom left plot we show the utility as a function of number of wavelets. Here utility is simply the ratio of the number of iterations spent in a model with a given number of wavelets to the total number of iterations in the RJMCMC run. In the bottom right, we plot the histogram of the normalized residuals (normalized by the TOA uncertainties) when including (blue) and not including (red) the MAP wavelet model. The green curve is a zero-mean unit variance Gaussian distribution.

First, we point out that although this ``event'' seems quite large, it is quite comparable to a real event in the published data of PSR B1855$+$09 \citep[see e.g. Figure 27 of][]{abb+15}. Next, we see that for both choices of amplitude prior we recover the injected waveform very well and recover the Gaussian nature of the underlying white noise shown in the bottom right panels. Most important, however, is the number of wavelets that the data prefer under the two different amplitude priors and how this effects the uncertainty in the recovered waveform. In the case where we have used a uniform prior on the SNR of the wavelet we see from the bottom left panel of Figure \ref{fig:bayeswavesnr} that the data prefer 6 wavelets and can support up to 11 wavelets, albeit with a much lower utility (utility is proportional to the Bayesian evidence). Furthermore we see from the wavelet model uncertainty (marginalized over all wavelet models) in gray in the top panel of the figure that the wavelet model only contributes to modeling the white noise burst and not any other features in the data. 

 Alternatively, for the log-uniform amplitude prior we see that the data prefer $\sim 18$ wavelets but can support $> 30$ resulting in very broad spread in the number of wavelets that are allowed by the data. Furthermore, we see from the waveform uncertainty at early times in the reconstructed waveform (gray shaded area of the top panel of Figure \ref{fig:bayeswavelog}) that many wavelets are being placed far away from the white noise burst. The fact that this prior allows for many more wavelets that are effectively sitting below the white noise level can be understood by the fact that there is quite a large prior volume at low amplitudes since a uniform prior in $\log_{10}A$ is proportional to a prior on the amplitude of $p(A)\propto A^{-1}$. In contrast, the uniform SNR prior is much more similar to a uniform amplitude prior. Thus, a-priori the log-uniform prior prefers low amplitude wavelets whereas the uniform SNR prior prefers high amplitude wavelets. 
 
 As mentioned above we want to choose a prior on the wavelet amplitude so that we use the minimum number of wavelets to model the signal. As we have seen from this example, a uniform SNR prior performs admirably in this respect whereas a log-uniform amplitude prior fails this test.
 
\begin{figure}
\includegraphics[scale=1.0]{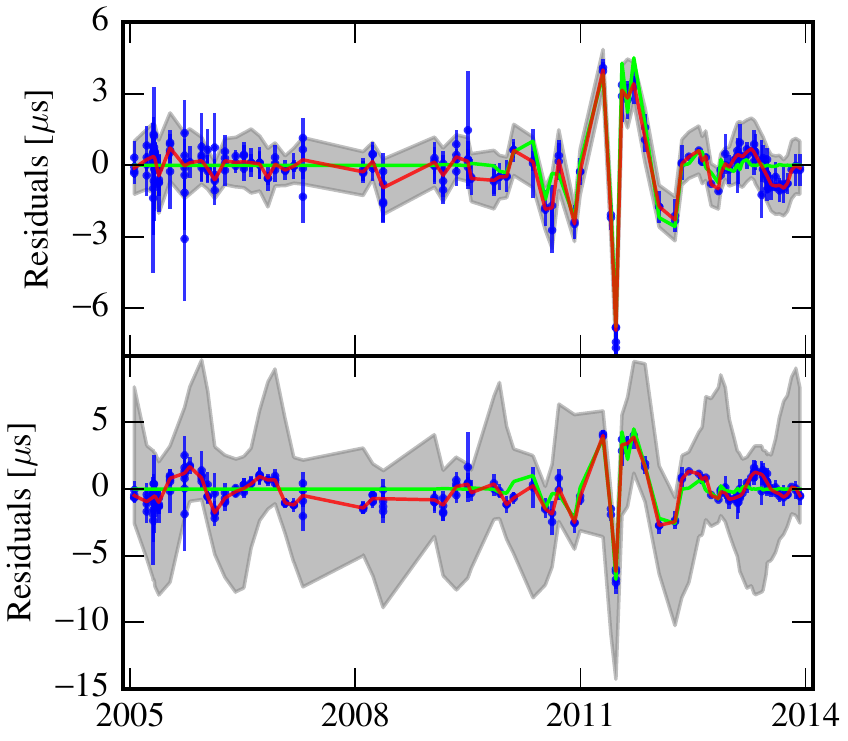} \\ 
\caption{Waveform reconstruction using a power-law (top) and free-spectrum (bottom) parameterization for the red noise with no additional non-stationary terms. The green is the injected waveform, the red is the MAP recovered waveform and the gray is the 90\% uncertainty on the waveform.}
\label{fig:standard} 
\end{figure}

 In Figure \ref{fig:standard}, we perform the analysis without the wavelet model using standard noise models with a power-law (top panel) and free-spectrum (bottom panel) red noise parameterization. 
In comparison with Figure \ref{fig:bayeswavesnr} we see that while the waveform is accurately recovered in both cases, the uncertainty in this inferred waveform is much larger than in our trans-dimensional wavelet model. Most importantly we note that the uncertainty is very large in the regions where the injected waveform is zero. This is mostly due to the fact that both the power-law and free-spectral models are constrained to be a realization of a stationary gaussian process. Furthermore we point out that the large uncertainty and periodic structure seen in the uncertainty region for the free-spectrum model is due to covariances between the power spectral amplitudes and the sky location and parallax terms in the timing model. This feature is unavoidable in such an unconstrained model. Finally, in these simulations we have only included white noise and the white noise burst, in such cases where there is steep red noise \emph{and} transient behavior, it is likely that standard methods would perform much worse.

\begin{figure}
\includegraphics[width=\columnwidth]{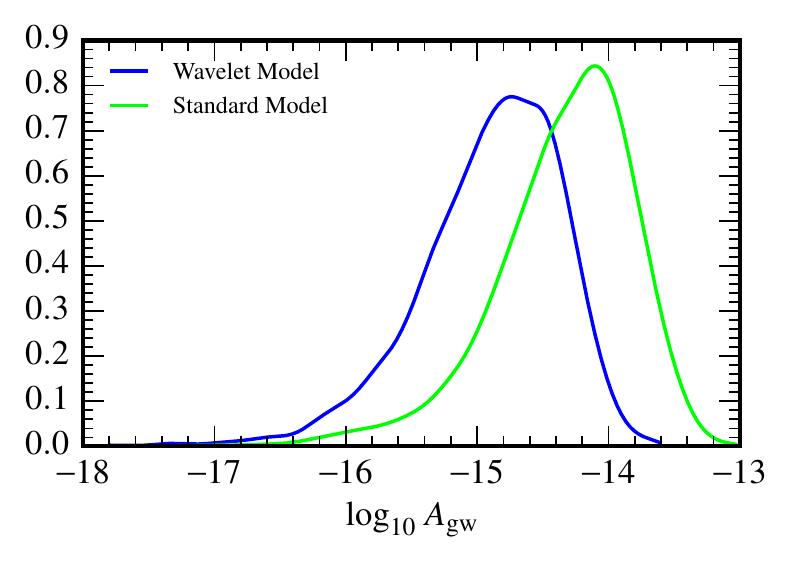} \\ 
\caption{Marginalized posterior distribution of the dimensionless strain amplitude of a stochastic GW background using the trans-dimensional wavelet model (blue) and standard power-law red noise model (green).}
\label{fig:gwb} 
\end{figure}

Lastly, we show that this kind of noise analysis could be crucial for placing tight constraints or making a detection of  a stochastic GW background. In Figure \ref{fig:gwb} we plot the marginalized posterior distribution of the dimensionless GW strain amplitude using the wavelet model above and the standard power-law red noise model, plotted in blue and green, respectively. For this simulation we use the same data that was analyzed above.  
In both cases we model the red noise via a power-law and include an additional noise term with a fixed spectral index (13/3) corresponding to a stochastic GW background of SMBHBs. For the wavelet model we allow up to 30 wavelets. As we see from Figure \ref{fig:gwb}, we can constrain the amplitude of a potential stochastic GW background significantly better when we include the additional wavelets in the noise model. Specifically, the upper 95\% upper limits on the GWB amplitude for the wavelet and non-wavelet model are $6.7\times 10^{-15}$ and $2.3\times 10^{-14}$, -- a factor of 3.5 improvement.  As our PTA data becomes more precise over time, this additional noise term could prove critical in either detecting or constraining the stochastic GW background.

\subsection{BayesSpecPTA}

Here we will test the adaptive spectral modeling techniques of Section \ref{sec:spectrum} on three cases that we call the null, intermediate, and extreme cases. For each simulation we will recover the spectrum using the adaptive technique, a standard power-law, and free spectral components. 

\subsubsection{Null Case}

\begin{figure}
\includegraphics[width=\columnwidth]{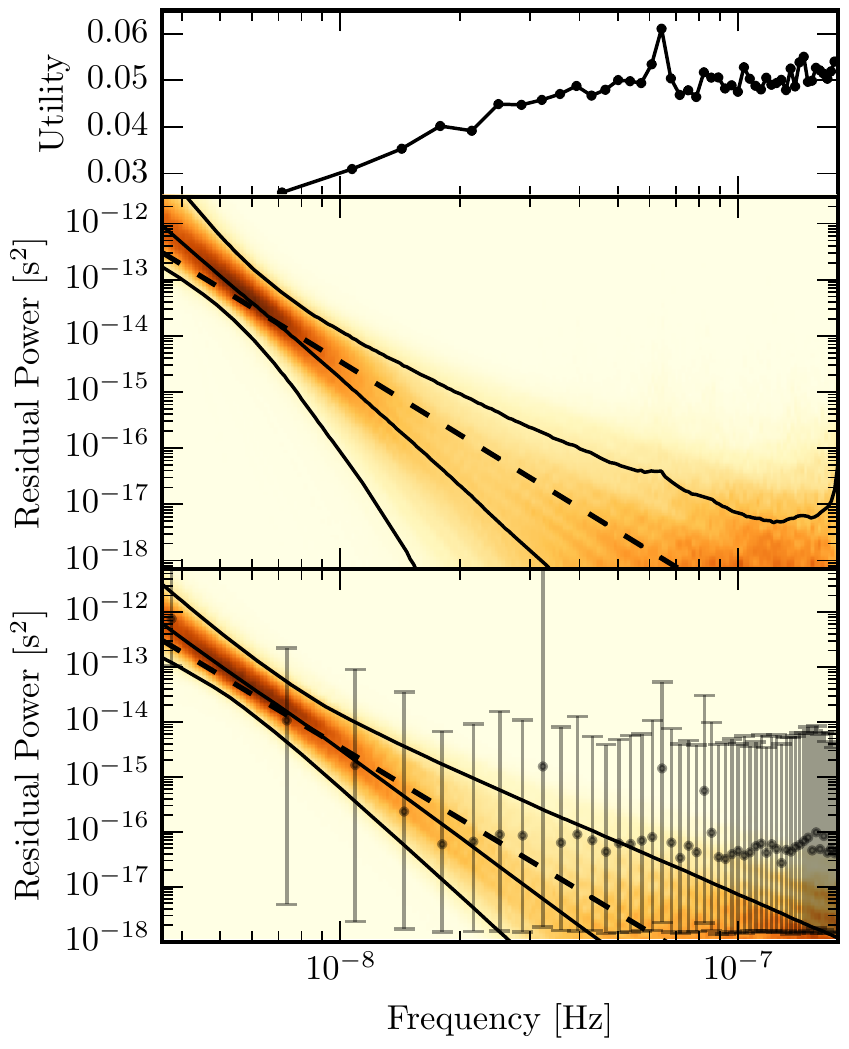} \\ 
\caption{Results of BayesSpecPTA and the two standard analyses (powerlaw and free spectral coefficients) on the null case. The top two panels show the recovered spectrum (middle panel) and the favored control points for the spectral interpolation (top panel), respectively. The bottom panel is the recovered spectrum from the powerlaw analysis (density) and the free spectral coefficients (error bars). In both density plots, the dashed line is the injected PSD and the solid black lines represent the median and 90\% confidence region. In the bottom panel, the gray error bars represent the median and 90\% confidence interval on each spectral coefficient.}
\label{fig:bayeslinespec0} 
\end{figure}

For the null case, we inject a realization of a gaussian process that follows a pure power-law distribution with a power spectral index of $13/3$ as one would expect from a gravitationally wave driven isotropic stochastic background from a population of SMBHBs in circular orbits. We have dubbed this the null case since the spectrum is fully described by two parameters and this is the simplest model that the adaptive spectral modeling technique can achieve. In Figure \ref{fig:bayeslinespec0} we show the results of our analysis on this null case. The middle panel shows the posterior probability density of the recovered spectrum from our adaptive method. The solid black lines are the 90\% credible intervals and the median value. The dashed line is the PSD of the injected noise process. The top panel shows the utility of the various control points as a function of frequency. Here we define the utility as the ratio of the number of iterations of the RJMCMC that a given control point was active to the total number of iterations. In this case we see that none of the additional control points was active for a significant number of iterations. The bottom panel again shows the posterior probability density of of the recovered spectrum using the power-law model.
\begin{figure}
\includegraphics[width=\columnwidth]{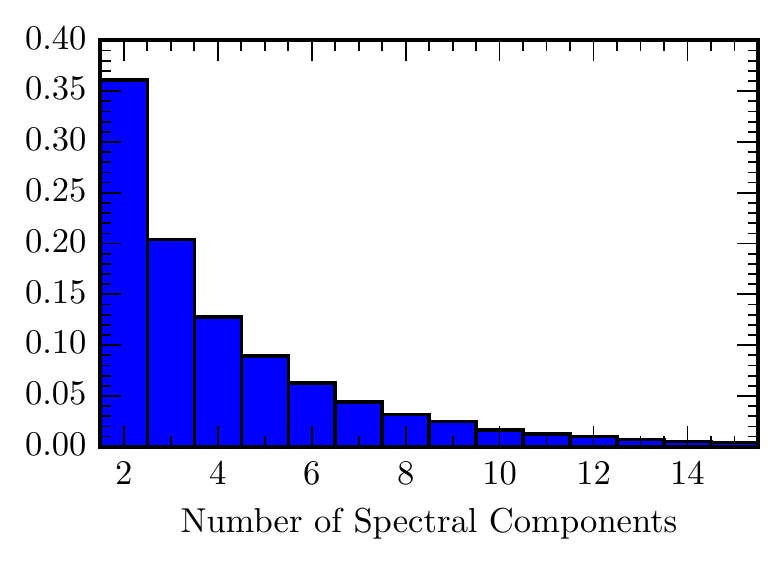} \\ 
\caption{Favored number of spectral components from the BayesSpecPTA analysis of the null case. In this case, where the injected spectrum is a simple powerlaw, the spectrum is best described by two parameters as one would expect.}
\label{fig:bayeslinehist0} 
\end{figure} 
 The gray points and error bars are the median and the 90\% confidence interval on each spectral component using the free spectral technique. From Figure \ref{fig:bayeslinehist0} we can see that in this case the adaptive technique does indeed favor no additional control points (i.e. the spectrum is parameterized by two parameters). Furthermore we see from Figure \ref{fig:bayeslinespec0} that the recovered spectrum is very similar for the adaptive and power-law models. The posterior is slightly broader in the adaptive case due to the fact that we allow for more than two spectral components and although the data favors only two, it is clear from Figure \ref{fig:bayeslinehist0} that the data does not heavily disfavor more than two components. Furthermore, the free spectral model is less constraining in the lowest frequency component and does not constrain other frequencies significantly. 

\subsubsection{Intermediate Case}

\begin{figure}[h]
\includegraphics[width=\columnwidth]{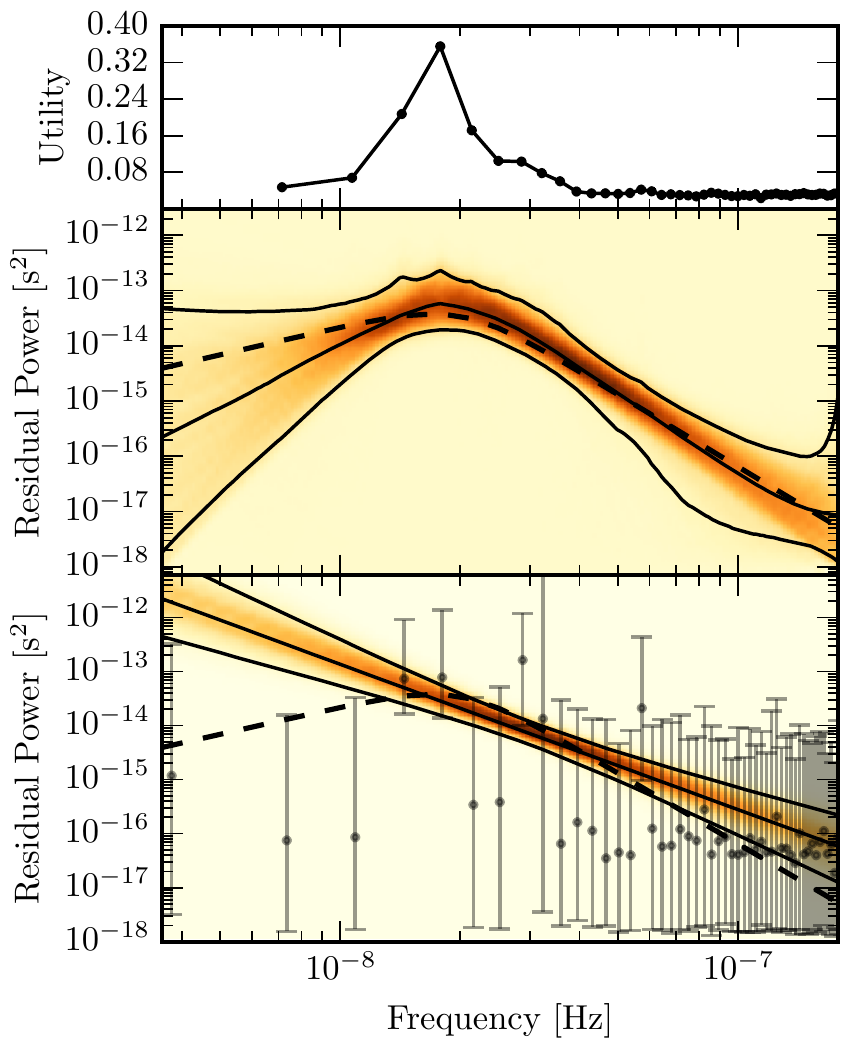} \\ 
\caption{Same as Fig. \ref{fig:bayeslinespec0} but for the intermediate case.}
\label{fig:bayeslinespec1} 
\end{figure}

\begin{figure}
\includegraphics[width=\columnwidth]{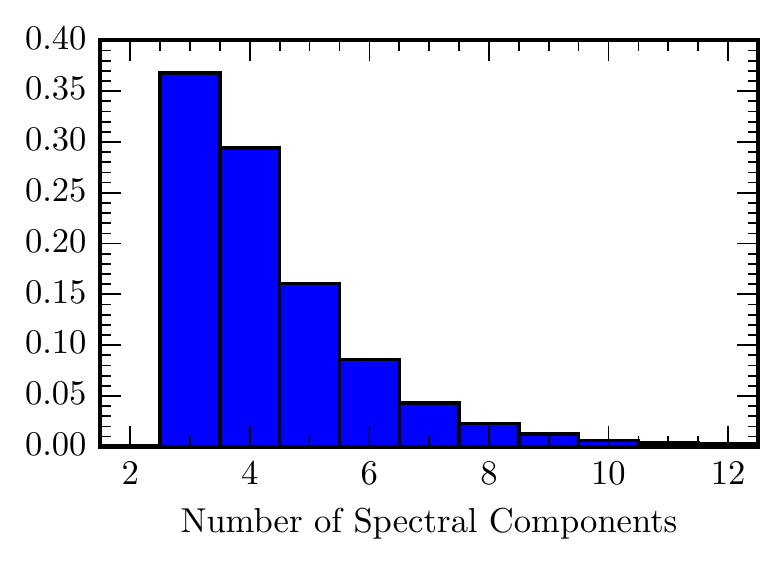} \\ 
\caption{Favored number of spectral components from the BayesSpecPTA analysis of the intermediate case. In this case, where the injected spectrum has a singe turnover the spectrum is best described by three parameters as one would expect, and two parameters (powerlaw) is strongly disfavored  while more than 3 components is allowed.}
\label{fig:bayeslinehist1} 
\end{figure}

For the intermediate case we have injected a realization of a gaussian process that follows a broken power-law distribution. In this case we see in Figure \ref{fig:bayeslinespec1} that the spectrum is well recovered with the adaptive technique and that the data strongly prefer an additional control point at the turnover frequency in the spectrum. Furthermore, we see from Figure \ref{fig:bayeslinehist1} that, overall, the data prefers to describe the spectrum by three parameters. This is exactly the behavior that we wish to see in that the most parsimonious model to describe a broken power-law contains three parameters. Meanwhile, the bottom panels of Figure \ref{fig:bayeslinespec1} shows that the power-law noise model significantly overestimates the low frequency noise and the free spectral model loosely models the power around the turnover in the spectrum. In fact, the powerlaw model does not even recover the spectrum within its 90\% credible interval, thus failing basic Bayesian consistency. This case really demonstrates the power of the adaptive technique in which it correctly identifies the most parsimonious model containing three parameters while still incorporating the uncertainty associated with more or less components, all while producing an accurate reconstruction of the power spectral density.

\subsubsection{Extreme Case}

\begin{figure}[h]
\includegraphics[width=\columnwidth]{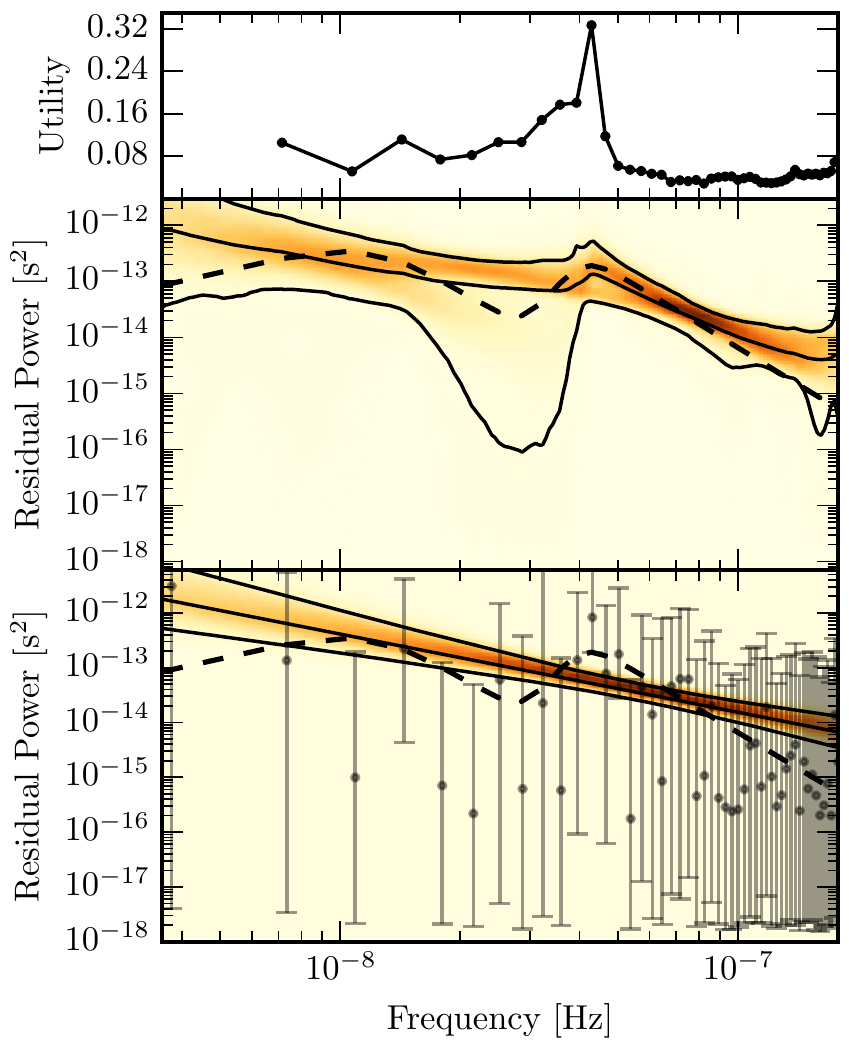} \\ 
\caption{Same content as Fig. \ref{fig:bayeslinespec0} for the extreme case.}
\label{fig:bayeslinespec2} 
\end{figure} 

\begin{figure}
\includegraphics[width=\columnwidth]{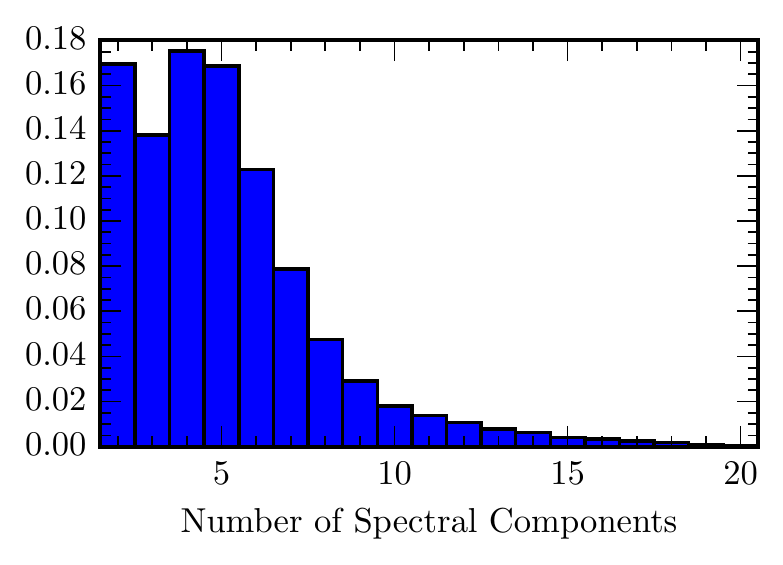} \\ 
\caption{Favored number of spectral components from the BayesSpecPTA analysis of the extreme case. In this case, the injected spectrum has two peaks. The data can either be described by a power law or a more complicated spectra that requires $\ge4$ control points.}
\label{fig:bayeslinehist2} 
\end{figure}

For the so-called extreme case we have injected a realization of a gaussian process that follows a distribution with two distinct peaks. In this case we see in Figure \ref{fig:bayeslinespec2} that the spectrum is well recovered with the adaptive technique and that the high frequency peak is clearly distinguishable while the low frequency peak is not as constrained. In this case, as can be seen in both Figures \ref{fig:bayeslinespec2} and \ref{fig:bayeslinehist2}, the model complexity is in strong competition with the goodness-of-fit in that data nearly equally prefer either a shallow power-law (2 parameters) that models the power at the peaks of the PSD but also the frequencies in between, or the more complex spectrum that models both the peaks \emph{and} valleys of the PSD. These features show the utility of this kind of analysis in that one does not have to a-priori choose a model for the spectrum and try to find the \emph{best} model from the data but instead we allow the data to constrain the model while \emph{marginalizing} over our uncertainty in that model.  For comparison, from the bottom panel of Figure \ref{fig:bayeslinespec2} we see that, because of the rigidness of the power-law model, it overestimates the power at low frequencies and underestimates the power at the high frequency peak. Furthermore, we see that even the free spectral method does not significantly constrain the PSD at either peak.

\subsection{Combination of Both Methods}

In this section we combine both the wavelet and adaptive signal modeling in one large RJMCMC where the number of wavelets and the number of control points are both free to vary. In principle, this method should be near optimal in that we allow the data to decide the complexity of the non-stationary and stationary noise processes simultaneously; however, in practice it can be difficult to distinguish between PSDs with significant high frequency power and transient signals. Therefore, in many cases, although we are able to recover the actual waveform in the data, it is difficult to separate out the stationary and non-stationary components. Nonetheless, because neither the wavelet modeling nor adaptive spectral modeling can optimally account for all noise sources alone, we recommend this combined approach.

\begin{figure}
\includegraphics[width=\columnwidth]{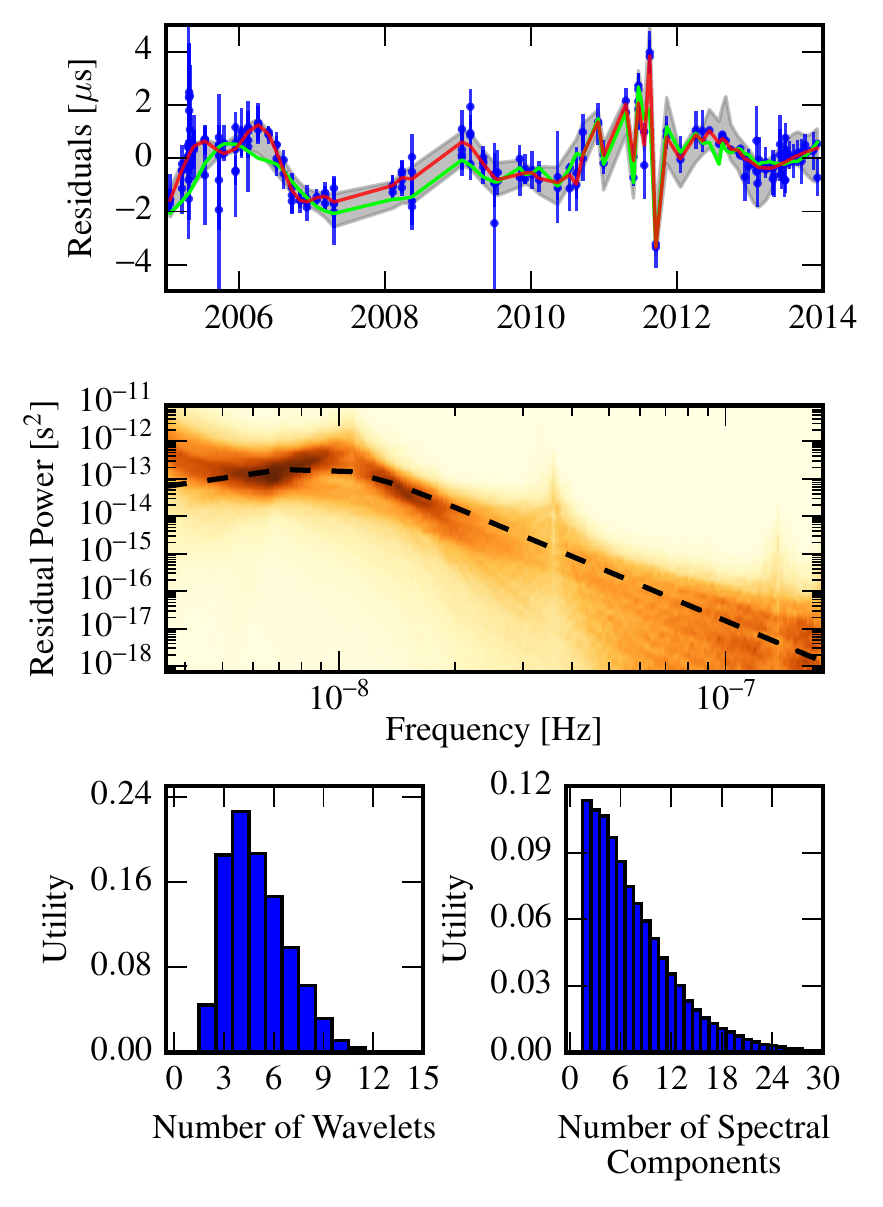} \\ 
\caption{Summary plots for combined run. In the top panel we plot the daily averaged residuals in blue, the MAP red noise and wavelet realization in red and the injected value in green. The gray region is the 90\% confidence region of the signal. In the middle panel we plot the probability density of the recovered PSD and the injected PSD in black. The bottom left panel shows the utility for the number of wavelets and number of spectral components.}
\label{fig:combine} 
\end{figure} 

To test this combined method we once again simulate a data set with low frequency red noise with a broken power spectrum (black dashed line in Figure \ref{fig:combine}) and a white noise burst. We have used a uniform SNR prior for the wavelet amplitudes and log-uniform prior for the amplitude of the interpolation control points. The results of this analysis are shown in Figure \ref{fig:combine}. In the top panel we plot the daily averaged residuals (blue) along with the injected (green) and MAP (red) red noise and transient noise signal realization and the corresponding 90\% credible region in gray. The middle panel shows the posterior probability of the recovered PSD along the injected PSD (dashed black line). The bottom panel shows the utility (same definition as above) of the number of wavelets and number of spectral components used in the model. 

First, we see that the model recovers the injected waveform very well through a combination of $\sim$ 5 wavelets and a broken power spectrum. Although this is a different realization of the same white noise burst used above the complexity of the wavelet model is directly comparable to that of Figure \ref{fig:bayeswavesnr}. Furthermore, we see the spectrum is recovered fairly well but there is still a large spread in which many spectral components could be used. This is due to the covariances between the wavelet model and the spectral model and is likely unavoidable unless we have some a-prior knowledge of the transient and/or the stationary noise spectrum.

\section{Discussion and Conclusions}
\label{sec:conclude}

As the precision of PTAs increases, sophisticated and robust noise modeling techniques are quickly becoming one of the most important aspects in the pulsar timing process. In this paper we have presented a trans-dimensional noise modeling approach that lets the data decide the complexity of the model as opposed to traditional methods of choosing a specific model prior to analysis. Furthermore, this analysis does not seek to determine the ``best'' model but instead captures our inherent uncertainty in the noise model by marginalizing over a large range of models. In this first analysis we have modified the transient signal modeling and adaptive spectral techniques first developed in the LIGO context \citep{cl15, lc15} for use in the PTA regime. To model non-stationary transient noise events we use a sum of Morlet-Gabor wavelets and to model our stationary red noise PSD we use a set of control points for linear interpolation on a fixed frequency grid. In both cases, the number of wavelets and control points are free to vary and final inferences are made by marginalizing over the full trans-dimensional model. Through simulations, we have shown that these methods perform better than standard methods both when applied separately and together. Furthermore we have shown that in the presence of strong non-stationary noise, the implementation of these trans-dimensional techniques could result in significantly more constraining upper limits on a stochastic GW background.

While the application in this work has been on achromatic (with to respect to radio frequency) single pulsar analysis, these methods could also be applied to dispersion measure (DM) modeling techniques and other radio-frequency dependent noise sources. Furthermore, this type of analysis could also be applied to multi-pulsar GW analysis both for noise modeling and for GW detection of unmodeled signals or as a cross check to other modeled GW searches.

\acknowledgments
JAE acknowledges support by NASA through Einstein Fellowship grant PF4-150120.
NJC was supported by NSF Physics Frontiers Center Award PFC-1430284.
Portions of this research were carried out at the Jet Propulsion Laboratory, California Institute of Technology, under a contract with the National Aeronautics and Space Administration. This work was supported in part by National Science Foundation Grant No. PHYS-1066293 and by the hospitality of the Aspen Center for Physics where this work was initiated.

\bibliography{bib,apjjabb}

\end{document}